\documentstyle{article}

\newcommand{\alphafive}{\alpha^{\scriptscriptstyle (5)}}
\newcommand{\alphathree}{\alpha^{\scriptscriptstyle (3)}}
\newcommand{\alphatwo}{\alpha^{\scriptscriptstyle (2)}}
\newcommand{\alphazero}{\alpha^{\scriptscriptstyle (0)}}
\newcommand{\alphanf}{\alpha^{\left(n_f\right)}}
\newcommand{\GeV}{{\rm GeV}}
\newcommand{\msbar}{{\scriptscriptstyle {\rm \overline{MS}}}}
\newcommand{\V}{{\scriptscriptstyle  V}}
\newcommand{\ainv}{{a^{-1}}}
\newcommand{\lat}{{\rm lat}}
\newcommand{\be}{\begin{equation}}
\newcommand{\ee}{\end{equation}}
\newcommand{\eq}[1]{Eq.~(\ref{#1})}
\newcommand{\order}{{\cal O}}
\newcommand{\PS}{\chi_b - \Upsilon}
\newcommand{\SS}{\Upsilon^\prime - \Upsilon}
\newcommand{\qchi}{q_{\chi_b}}
\newcommand{\qups}{q_{\Upsilon^\prime}}

\begin{document}


\title{ A Precise Determination of $\alpha_s$ From Lattice QCD}

\author{
C.~T.~H.~Davies,$^a$ K.~Hornbostel,$^b$
G.~P.~Lepage,$^c$\\ A.~Lidsey,$^a$
J.~Shigemitsu,$^d$ J.~Sloan,$^e$\\[.4cm]
\small $^a$University of Glasgow, Glasgow, UK G12 8QQ \\
\small $^b$Southern Methodist University, Dallas, TX 75275 \\
\small $^c$Newman Laboratory of Nuclear Studies,
            Cornell University, Ithaca, NY 14853 \\
\small $^d$The Ohio State University, Columbus, OH 43210 \\
\small $^e$Florida State University, SCRI, Tallahassee, FL 32306 \\
\\ }

\date{August 1994}

\maketitle

\begin{abstract}
We present a new determination of the QCD strong coupling
constant based on precise lattice calculations of the
$\Upsilon$ spectrum.  The largest systematic uncertainty in previous
such determinations resulted from the absence of vacuum polarization
from light quarks.  We substantially reduce this error by
including two flavors of dynamical light quarks and
extrapolating to three.  We find
$\alphathree_\V(8.2~\GeV) = 0.196(3)$ for three light
flavors, corresponding to $\alphafive_\msbar(M_Z) = 0.115(2)$.
This is significantly more accurate than previous determinations
using this or any other technique.
\end{abstract}

In recent years it has become apparent that numerical simulations of lattice
quantum chromodynamics (QCD) could provide accurate and reliable determinations
of the strong interaction coupling
constant~\cite{aida,latconf-japan,onogi,latconf-dallas,luscher-schilling}.
The largest systematic
error in previous lattice determinations resulted from the neglect of quark
vacuum polarization, whose sizable
contribution was estimated perturbatively. In this paper
we significantly reduce this systematic error by presenting simulation
results that include dynamical light quarks.
We find $\alphathree_\V(8.2~\GeV) = 0.196(3)$ for the
strong coupling constant with~$n_f=3$ flavors of
light quarks; $V$ denotes the physical scheme based
on the static-quark potential, discussed in
Refs.~\cite{blm,lm}. The corresponding
${\rm \overline{MS}}$~coupling at the $Z$~mass is
$\alphafive_\msbar(M_Z) = 0.115(2)$,
consistent with the world average of
0.117(5)~\cite{worldave}, but significantly more accurate.

In this study we used two different sets of simulations, one without quark
vacuum polarization \mbox{$(n_f=0)$},
the other with two flavors of light quarks
\mbox{$(n_f=2)$}~\cite{gregurs}.
The $n_f=2$ gauge-field configurations were generated using
a staggered-quark action for light quarks with the
Hybrid Molecular Dynamics algorithm,
and both used the standard Wilson action for gluons.
For $b$ quarks, we employed the
nonrelativistic formulation of quark dynamics (NRQCD);
details are in Refs.~\cite{spect,mb}.

There are two steps in a determination of the coupling constant from
lattice simulations.
The first is to specify or accurately determine the parameters of the
lattice lagrangian, and in particular the lattice spacing~$a$.
The second is to use this lagrangian to compute nonperturbatively
an appropriate short-distance quantity.  Comparison with the
perturbative expansion for the same quantity fixes the coupling.

Just as for continuum QCD, the bare coupling constant and masses
must be provided as input.  In the lattice action, the bare
coupling $g_\lat$ appears in the parameter $\beta = 6/g^2_\lat$.
The lattice spacing is not an input.
Rather, it is specified implicitly by $\beta$; for each
$\beta$ there is a corresponding $a$.
By expressing all dimensionful quantities
in units of~$a$, it is scaled out of the
action, and so serves to set the overall mass scale.
As a result,
the simulation produces a value for some particular mass $M$
only in the dimensionless combination $a M$.
We must therefore know $a$ before we can compare $M$ to its
experimental value.
In our simulation, we compute the mass difference
between the~$\Upsilon$ and
$\Upsilon^\prime$ mesons, $a\Delta M(\SS)$,
and between the $\Upsilon$ and the
spin average of the $\chi_b$ states, $a\Delta M(\PS)$.
We then divide these by the
experimentally measured mass differences to obtain two
independent estimates for $a$.

Heavy-quark systems possess several properties which permit
us to measure $a$ accurately~\cite{latconf-japan}.
They are essentially nonrelativistic;
the use of a nonrelativistic effective action to exploit this allows
a large portion of the spectrum to be computed efficiently and
precisely.  They are physically small,
and do not suffer from finite-volume errors on modestly sized
lattices.  Their spin-averaged mass splittings are observed experimentally
to be nearly independent of the heavy-quark mass,
varying by only a few per cent between the $\Upsilon$ and $\psi$,
making our results insensitive to tuning errors in the bare $b$-quark mass.
Because including vacuum polarization in simulations from nearly massless
quarks
is difficult, it is common to use unrealistically large
$u$- and $d$-quark masses, and then extrapolate to the correct values.
However, due to the small size of the $\Upsilon$ and the large
momentum transfers between its constituents,
the bare light-quark masses of about 25 MeV
in our $n_f=2$ simulations are negligible,
making extrapolation unnecessary.
Finally, $\Upsilon$ decay rates are negligible as compared to their
energy splittings, making the effect of light-quark mass values on
nonanalytic threshold behavior unimportant.

Our results for $n_f=0$ and $2$ are summarized in Table~\ref{splittings}.
The $b$-quarks in $\Upsilon, \Upsilon^\prime$ and $\chi_b$ mesons
typically exchange momenta of order $1~\GeV$,
so that the appropriate number of light flavors
to include in a study of their dynamics is three.
Having results for $n_f$ of both $0$ and $2$ will allow us to
accurately extrapolate to \mbox{$n_f=3$}.  In addition, we varied
the bare $b$-quark mass around the correct value of
\mbox{$a\,M_b^0=1.7(1)$}.  As expected, the splittings showed little
sensitivity to $M_b^0$.
\begin{table}
\begin{center}
\begin{tabular}{ccc|cc|ll}
$\beta$ & $n_f$ & $a\,M_b^0$ & \multicolumn{2}{c|}{$\SS$}
 & \multicolumn{2}{c}{$\PS$} \\
&&& $a\,\Delta M$ & $\ainv$ & $a\,\Delta M$ & $\ainv$ \\ \hline
6.0 & 0 & 1.71  &  .241(11) & 2.34(11)  &  .171(8) & 2.57(12) \\
    &&    1.80  &  .239(11) & 2.36(11)  &  .174(12)& 2.53(18) \\
    &&    2.00  &  .235(11) & 2.40(11)  &  .173(10)& 2.54(15) \\
5.6 & 2 & 1.80  &  .237(10) & 2.38(10)  &  .178(5) & 2.47(7)

\end{tabular}
\end{center}
\caption{Lattice QCD simulation results for the difference between the
$\Upsilon^\prime$ and $\Upsilon$ masses, and between the spin-averaged
$\chi_b$~mass and the $\Upsilon$~mass. Results are given for different bare
gluon couplings~$\beta = 6/g^2_\lat$ and bare quark masses~$M_b^0$, and for
$n_f=0$ and~2 flavors of light quarks. The correct bare mass for the
$b$-quark is $a\,M_b^0 = 1.7(1)$. The splittings are
corrected for $\order(a^2)$ errors in the gluon action.
The errors shown are statistical
and result from our use of Monte Carlo methods in the simulations. Values for
$\ainv$ are in GeV and are obtained using
$\Delta M(\Upsilon^\prime - \Upsilon) = 0.563~\GeV$ and $\Delta
M(\chi_b-\Upsilon) = 0.440~\GeV$.}
\label{splittings}
\end{table}

\begin{figure}
\begin{center}
\setlength{\unitlength}{.025in}
\begin{picture}(130,120)(10,930)
\put(15,940){\line(0,1){120}}
\multiput(13,950)(0,50){3}{\line(1,0){4}}
\multiput(14,950)(0,10){10}{\line(1,0){2}}
\put(12,950){\makebox(0,0)[r]{9.5}}
\put(12,1000){\makebox(0,0)[r]{10.0}}
\put(12,1050){\makebox(0,0)[r]{10.5}}
\put(12,1060){\makebox(0,0)[r]{GeV}}

\multiput(75,1062)(3,0){3}{\line(1,0){2}}
\put(84,1062){\makebox(0,0)[l]{ Experiment}}
\put(80,1056){\makebox(0,0)[tl]{\circle*{2}}}
\put(84,1056){\makebox(0,0)[l]{ $n_f = 0$}}
\put(80,1050){\makebox(0,0)[tl]{\circle{2}}}
\put(84,1050){\makebox(0,0)[l]{ $n_f = 2$}}

\put(27,940){\makebox(0,0)[t]{${^1S}_0$}}
\put(25,943){\circle*{2}}
\put(30,942){\circle{2}}

\put(52,940){\makebox(0,0)[t]{${^3S}_1$}}
\put(68,946){\makebox(0,0){1S}}
\multiput(43,946)(3,0){7}{\line(1,0){2}}
\put(50,946){\circle*{2}}
\put(55,946){\circle{2}}

\put(68,1002){\makebox(0,0){2S}}
\multiput(43,1002)(3,0){7}{\line(1,0){2}}
\put(50,1004){\circle*{2}}
\put(50,1004){\line(0,1){3}}
\put(50,1004){\line(0,-1){3}}
\put(55,1003){\circle{2}}
\put(55,1004){\line(0,1){2}}
\put(55,1002){\line(0,-1){2}}

\put(68,1036){\makebox(0,0){3S}}
\multiput(43,1036)(3,0){7}{\line(1,0){2}}
\put(50,1034){\circle*{2}}
\put(50,1034){\line(0,1){12}}
\put(50,1034){\line(0,-1){12}}
\put(55,1029){\circle{2}}
\put(55,1030){\line(0,1){13}}
\put(55,1028){\line(0,-1){13}}

\put(92,940){\makebox(0,0)[t]{${^1P}_1$}}

\put(108,990){\makebox(0,0){1P}}
\multiput(83,990)(3,0){7}{\line(1,0){2}}
\put(90,987){\circle*{2}}
\put(90,987){\line(0,1){2}}
\put(90,987){\line(0,-1){2}}
\put(95,990){\circle{2}}

\put(108,1026){\makebox(0,0){2P}}
\multiput(83,1026)(3,0){7}{\line(1,0){2}}
\put(90,1032){\circle*{2}}
\put(90,1032){\line(0,1){7}}
\put(90,1032){\line(0,-1){7}}
\put(95,1025){\circle{2}}
\put(95,1026){\line(0,1){4}}
\put(95,1024){\line(0,-1){4}}

\put(127,1020){\makebox(0,0){1D}}
\put(120,940){\makebox(0,0)[t]{${^1D}_2$}}
\put(120,1020){\circle*{2}}
\put(120,1020){\line(0,1){7}}
\put(120,1020){\line(0,-1){7}}

\end{picture}
\end{center}

 \caption{NRQCD simulation results for the spectrum of the
$\Upsilon$ system, including radial excitations.
  Dashed lines indicate experimental values for the triplet
$S$-states, and for the
 spin-average of the triplet $P$-states. The energy zero from
 simulation results is adjusted to give the correct mass to the
 $\Upsilon(1{^3S}_1)$. Results are from a simulation with $n_f=0$
(filled circles) and from one with $n_f=2$ (open circles),
using $\ainv=2.4~\GeV$ for both.  The errors shown are statistical;
systematic errors are several tens of MeV.}
\label{spect-fig}
\end{figure}
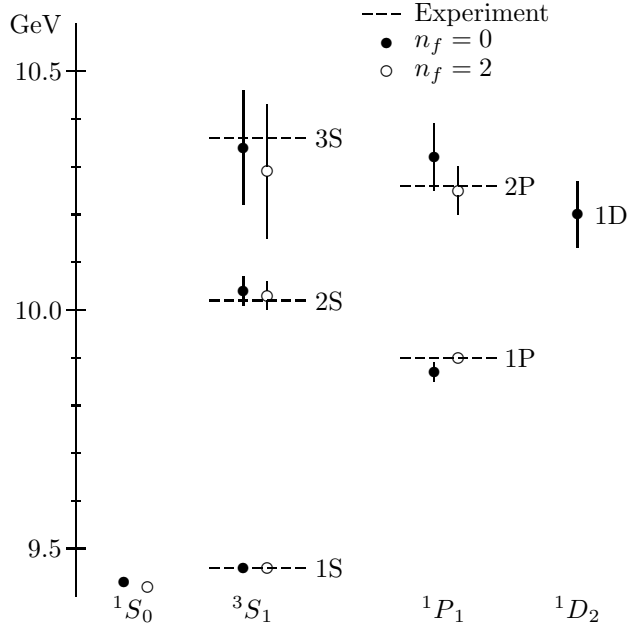

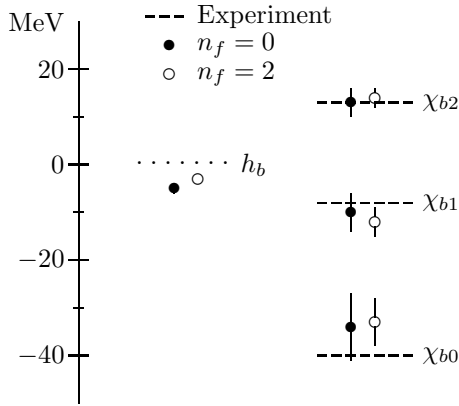
\begin{figure}
\begin{center}
\setlength{\unitlength}{.025in}
\begin{picture}(100,80)(5,-50)
\put(15,-50){\line(0,1){80}}
\multiput(13,-40)(0,20){4}{\line(1,0){4}}
\multiput(14,-40)(0,10){7}{\line(1,0){2}}
\put(12,-40){\makebox(0,0)[r]{$-40$}}
\put(12,-20){\makebox(0,0)[r]{$-20$}}
\put(12,0){\makebox(0,0)[r]{$0$}}
\put(12,20){\makebox(0,0)[r]{$20$}}
\put(12,30){\makebox(0,0)[r]{MeV}}

\multiput(29,31)(3,0){3}{\line(1,0){2}}
\put(38,31){\makebox(0,0)[l]{ Experiment}}
\put(34,25){\makebox(0,0)[tl]{\circle*{2}}}
\put(38,25){\makebox(0,0)[l]{ $n_f = 0$}}
\put(34,19){\makebox(0,0)[tl]{\circle{2}}}
\put(38,19){\makebox(0,0)[l]{ $n_f = 2$}}



\multiput(27,0)(3,0){7}{.}
\put(49,0){\makebox(0,0)[l]{$h_b$}}
\put(35,-5){\circle*{2}}
\put(35,-5){\line(0,1){1}}
\put(35,-5){\line(0,-1){1}}
\put(40,-3){\circle{2}}

\multiput(65,-40)(3,0){7}{\line(1,0){2}}
\put(87,-40){\makebox(0,0)[l]{$\chi_{b0}$}}
\put(72,-34){\circle*{2}}
\put(72,-34){\line(0,1){7}}
\put(72,-34){\line(0,-1){7}}
\put(77,-33){\circle{2}}
\put(77,-32){\line(0,1){4}}
\put(77,-34){\line(0,-1){4}}

\multiput(65,-8)(3,0){7}{\line(1,0){2}}
\put(87,-8){\makebox(0,0)[l]{$\chi_{b1}$}}
\put(72,-10){\circle*{2}}
\put(72,-10){\line(0,1){4}}
\put(72,-10){\line(0,-1){4}}
\put(77,-12){\circle{2}}
\put(77,-11){\line(0,1){2}}
\put(77,-13){\line(0,-1){2}}

\multiput(65,13)(3,0){7}{\line(1,0){2}}
\put(87,13){\makebox(0,0)[l]{$\chi_{b2}$}}
\put(72,13){\circle*{2}}
\put(72,13){\line(0,1){3}}
\put(72,13){\line(0,-1){3}}
\put(77,14){\circle{2}}
\put(77,15){\line(0,1){1}}
\put(77,13){\line(0,-1){1}}

\end{picture}
\end{center}
 \caption{NRQCD simulation results for the spin structure of the
  lowest-lying $P$-states.
  Dashed lines indicate experimental values for the triplet
$P$-states; the dotted line shows their spin average.
Masses are relative
to the spin-averaged state. Results are from a simulation with $n_f=0$
(filled circles) and from one with $n_f=2$ (open circles),
using $\ainv=2.4~\GeV$ for both.
The errors shown are statistical; systematic errors are within about
5~MeV.}
\label{spect-splits}
\end{figure}

We have determined~$\ainv$ both by fitting
the splittings separately and simultaneously.
Fitting separately produced a discrepancy in~$\ainv$ of a couple standard
deviations for $n_f=0$, and about half this for $n_f=2$.
As we will show, this small discrepancy vanishes when we extrapolate
$n_f$ to three.  Insofar as the effect is real, it is likely due to the
larger intrinsic momentum transfers for
$S$ states as opposed to $P$ states.\footnote{
Perturbation theory, though not justified at these momenta,
provides a qualitative explanation of the effect on the determination
of $\ainv$.
The centrifugal barrier makes the average separation
between the quarks in the $P$-state $\chi_b$ larger than for the $S$-state
$\Upsilon$ or $\Upsilon^\prime$, as is familiar from hydrogen or positronium.
Consequently, the
typical exchanged momentum for $\chi_b$ quarks, $\qchi$, is smaller
than $\qups$.  The perturbative binding energy is given by
$\alpha^2_\V(q) C_F^2 M_b/16$, with $q=\qups$ for
$\Upsilon^\prime$ and $\qchi$ for $\chi_b$.
Since $\qchi < \qups$, the $\chi_b$ is more tightly bound.
However, for $n_f=0$, this effect is exaggerated, as $\alphazero_\V(q)$
increases more quickly than $\alphathree_\V(q)$ with decreasing $q$.
Thus, for $n_f<3$, $\Delta M(\PS)$ should be underestimated relative
to $\Delta M(\SS)$, as is observed.  Fitting to data would then require a
larger $\ainv$ for $\Delta M(\PS)$ than for $\Delta M(\SS)$.
}
Fitting simultaneously, we find that our simulation data are
consistent with $\ainv\approx 2.4~\GeV$.
Our results for other low-lying excitations and spin splittings
of the $\Upsilon$ system, using this value for $\ainv$, are
displayed in Figures~\ref{spect-fig} and~\ref{spect-splits},
where they are compared with their
experimental values.
The excellent agreement supports the reliability
of our simulations. We emphasize that these are calculations from
first principles; our approximations can be systematically improved.
The only inputs are the lagrangians describing gluons and quarks, and the
only parameters are the bare coupling constant and quark mass. In particular,
these simulations are {\em not\/} based on a phenomenological quark
potential model.

The $b$-quark action includes relativistic corrections in the quark
velocity $v$ to $\order(v^2)$, and corrections
to $\order(a^2)$ in the lattice spacing.
The leading systematic error affecting~$\ainv$ is almost certainly
due to finite-lattice-spacing errors in the gluon action.
To correct for this requires an additional~$\order(a^2)$ interaction.
As it is sensitive to short distances,
the effect of this interaction can be estimated using perturbation theory,
which gives a mass shift of
\be
a\Delta M_g = \frac{4\pi\,\alpha_\V(q_\delta)}{15}\,a^3\,\left| \psi(0)
\right|^2 \, .
 \ee
Here $q_\delta\approx 5~\GeV$ is the typical momentum transferred
through the interaction, and $\psi(0)$ is the meson wavefunction evaluated
at the origin. Using $n_f = 0$ simulation results for
$\psi(r)$ and $\alpha_\V$~\cite{spect}, we find that $a\Delta
M_g$ is 0.0036 for the $\Upsilon$ and 0.0023 for the $\Upsilon^\prime$
when $aM_b^0=1.71$; there is no shift for the $P$-state $\chi$'s. The
correction increases approximately linearly with mass and is about
30\% larger for $n_f=2$.
The splittings in Table~\ref{splittings} include this correction
(although Figures~\ref{spect-fig} and~\ref{spect-splits} do not).
It shifts $a^{-1}$ by almost a standard
deviation in the case of the $\chi_b-\Upsilon$ splittings, but is
negligible for the $\Upsilon^\prime-\Upsilon$ splittings.

To check the validity of Eq.~(1), we examined the perturbative
prediction for the hyperfine splitting between the $\Upsilon$ and
$\eta_b$, given by the similar formula
\be
\Delta M_{\rm hfs} =
\frac{32\pi\,\alpha_\V(q_\delta)}{9\,M_b^2}\,\left|\psi(0)\right|^2 ,
\ee
where $M_b$ is the pole mass of the $b$-quark~\cite{mb}. This formula
gives $a\,\Delta M_{\rm hfs} = 0.0122(10)$ when $a\,M_b^0=1.71$ and
$n_f=0$, which compares very well with the nonperturbative result
0.0123(2) we obtain from our simulation~\cite{spect}. This suggests
that our perturbative estimates of the $\order(a^2)$ shifts (Eq.~(1))
are quite reliable.
We have also computed these shifts nonperturbatively
using a lattice potential model, and found essentially identical
corrections.
Other systematic errors are higher order
in $a$ or in the quark velocity $v$, and so are most
likely negligible. The dominant errors in determining $\ainv$ are
statistical.

Having determined the lattice spacing, the second step
is to use simulation results to extract the coupling constant.
Our approach closely parallels determinations based on high-energy
phenomenology, with the simulation playing the role of
an experimental measurement.
We identify short-distance quantities whose perturbative
expansions are known at least through second order, and then determine these
quantities nonperturbatively using our simulation. By equating the
perturbative expansion to the nonperturbative value, we can solve for the
strong coupling constant.

One of the most ultraviolet quantities in lattice QCD is the expectation
value of the $1\times1$~Wilson loop operator. Its perturbative expansion
to $\order(\alpha_\V^2)$ is~\cite{weisz,hk,bitar,lm}
 \be \label{plaq}
 -\ln W_{1,1} = \frac{4\pi}{3}\,\alphanf_\V(3.41/a)\,\left\{ 1 -
 \left( 1.185 + 0.070\,n_f\right) \,\alphanf_\V \right\}.
 \ee
The strong coupling constant~$\alpha_\V$ was defined in Refs.~\cite{blm,lm}
in terms of the static-quark potential.
To simplify the discussion of larger Wilson loops that follows,
it is convenient to regard \eq{plaq},
with no higher-order terms in~$\alpha_\V$,
as defining~$\alpha_\V$.
The two definitions differ only at $\order(\alpha_\V^3)$.
The scale~$3.41/a$ follows from the technique
described in Ref.~\cite{lm}, and indicates the important
momentum scale in~$W_{1,1}$.  It corresponds to 8 - 9~GeV for our lattices,
confirming that $W_{1,1}$ is very ultraviolet.

\begin{table}
\begin{center}
\begin{tabular}{c|rrrcc|cc}
 \multicolumn{8}{c}{$\beta = 6.0 \;\;\; n_f = 0$ \hspace{25pt} }  \\
 loop & $c_1$ & $c_2$ & $c_3$ & $a\,q_{n,m}$ &
    sim'n & \multicolumn{2}{c}{$\alphanf_\V(3.41/a)$} \\
    &&&&&& $\order(\alpha^2)$ & $\order(\alpha^3)$  \\ \hline\hline
 $-\ln W_{1,1}$ & 4.19 & $-4.96$ & 0 & 3.41 & 0.5214(0) & .152 & .1517 \\
 $-\ln W_{1,2}$ & 7.22 & $-7.57$ & 2.6 & 3.07 & 0.9582(1) & .154 & .1522 \\
 $-\ln W_{1,3}$ & 10.07 & $-9.60$ & 5.3 & 3.01 & 1.3757(2) & .155 & .1525 \\
 $-\ln W_{2,2}$ & 11.47 & $-10.58$ & 11.1 & 2.65 & 1.6605(3) & .158 & .1532
\\[.35cm]
  & \multicolumn{5}{|c|}{$\beta = 5.6 \;\;\; n_f = 2 \;\;$} & & \\[.15cm]
\hline\hline
 $-\ln W_{1,1}$ & 4.19 & $-5.55$ & 0 & 3.41 & 0.5708(1) & .179 & .1785 \\
 $-\ln W_{1,2}$ & 7.22 & $-8.51$ & -- & 3.07 & 1.0522(1) & .181 & -- \\
 $-\ln W_{1,3}$ & 10.07 & $-10.89$ & -- & 3.01 & 1.5123(2) & .181 & -- \\
 $-\ln W_{2,2}$ & 11.47 & $-11.84$ & -- & 2.65 & 1.8337(3) & .185 & --
\end{tabular}
\end{center}
\caption{Perturbative and simulation results for several small Wilson
loops~[15].  The entries in the final two columns list the values for
$\alpha_\V$ extracted by comparing the simulation results for each
$-\ln W$ to its perturbative expansion, correct to second and third order,
respectively.}
\label{lnw-table}
\end{table}

Simulation results for $-\ln W_{1,1}$ are listed in Table~\ref{lnw-table}.
{}From \eq{plaq} and the $\ainv$ extracted from the $\PS$ splitting we find
\be
\alphazero_\V(8.76(41)~\GeV) = 0.1517\, ,
\ee
\be
\alphatwo_\V(8.42(24)~\GeV) = 0.1785
 \ee
or equivalently,
\be
\alphanf_\V(8.2~\GeV) = \cases{
 0.1548(23) & \mbox{for $n_f=0$} \cr
 0.1800(16) & \mbox{for $n_f=2$}\, , \cr }
\ee
where the errors are due to the statistical errors in~$\ainv$. As mentioned
above, we must
extrapolate our results to~$n_f=3$. Perturbation theory suggests that
$1/\alphanf_\V$ is more nearly linear  for small changes in $n_f$ than
$\alphanf_\V$, and so we extrapolate the inverse couplings to obtain
 \be
 \alphathree_\V(8.2~\GeV) = 0.1959(34)\, . \;\;\;\;\;\;\;  \{\PS\}
 \ee
Since we are extrapolating by only 9\%, extrapolation errors are probably
smaller than the statistical errors quoted here;
perturbation theory indicates that they are about 0.2\%, which is
negligible.

Repeating this analysis using $\ainv$ from $\SS$ gives
\be
\alphanf_\V(8.2~\GeV) = \cases{
 0.1504(22) & \mbox{for $n_f=0$} \cr
 0.1779(23) & \mbox{for $n_f=2$}\, , \cr }
\ee
and
 \be
 \alphathree_\V(8.2~\GeV) = 0.1958(46)\, , \;\;\;\;\;\;\;  \{\SS\}
 \ee
which agrees with the $\PS$ value.
We have then,
from both $\PS$ and $\SS$, our
primary result:\footnote{We expect these two determinations
to be statistically correlated, and quote as an error the uncertainty in
the $\PS$ determination, rather than combining the errors as if independent.}
 \be \label{main-result}
 \alphathree_\V(8.2~\GeV) = 0.1959(34)\, .
 \ee

Because the internal momenta transferred by the $\Upsilon$ constituents
are small relative to $c$- and $b$-quark masses,
it would be incorrect to extrapolate $\alpha_\V$ obtained with
$n_f=0$ and $2$ light quarks directly to $n_f = 4$ or $5$.
The correct way to incorporate
these heavier flavors is to run the coupling to below the $c$-quark
threshold, then apply matching conditions
as $\alpha$ is run back up through the $c$ and $b$ thresholds.
We will use this procedure to obtain $\alphafive_\msbar$ at the $Z$ mass.

We chose the $1\times1$~Wilson loop because we expected nonperturbative
effects to be very small due to the large momentum scale it probes.
To verify this, we have determined $\alphanf_\V$ using larger
Wilson loops and loops with different shapes, which should
have significantly larger nonperturbative contributions.
In Table~\ref{lnw-table} we give the perturbative expansion coefficients
for the smallest Wilson loops.  These coefficients are defined by
\be
-\ln W_{n,m}^{(n_f)} = \sum_{i=1} c_i^{(n_f)}(n,m)\,
          \left[\alphanf_\V(q_{n,m})\right]^i \, ,
\ee
with $\alphanf_\V$ defined by \eq{plaq}~\cite{w-pert-th}.
We also quote
simulation results for these quantities, and values for $\alphanf_\V(3.41/a)$
obtained by matching second- and third-order perturbative expansions to the
simulation results. The $n_f=0$~results, when third-order
perturbation theory is used,
show that the small loops all give the same value for
$\alphazero_\V(3.41/a)$ to within less than one per cent.
This confirms that nonperturbative effects are
completely negligible.
(In fact, the slight variation is likely due to fourth-order
perturbative corrections.)
The $n_f=2$ results are consistent with this
conclusion, although the test is somewhat less stringent since the
third-order perturbative coefficients are not known for $n_f\ne0$.

To verify that additional lattice-spacing errors are under control,
in Table~\ref{scaling-table}
we compare our result for $\alphazero(8.2~\GeV)$ computed from a
simulation at $\beta=6.0$ with those from
$\beta = 6.2$ and $6.4$~\cite{catterall}.  The scales for the
corresponding $1 \times 1$ Wilson loops range from 8 to 14~GeV.  That
these give consistent values for $\alphazero(8.2~\GeV)$ indicates
that, within errors, the coupling constant is scaling correctly,
and confirms results found in Ref.~\cite{lm}.

\begin{table}
\begin{center}
\begin{tabular}{cc|ccc|c}
$\beta$ & $-\ln W_{1,1}$ & $a^{-1}$ & $q_\Box$ &
  $\alphazero_\V(q_\Box)$ & $\alphazero_\V(8.2~\GeV)$ \\ \hline\hline
6.4 & 0.4610 & 4.12(63) & 14(2) & 0.1302 & 0.151(7) \\
6.2 & 0.4884 & 3.50(33) & 12(1) & 0.1381 & 0.156(5) \\
6.0 & 0.5214 & 2.57(12) & 8.8(4)& 0.1517 & 0.155(2)
\end{tabular}
\end{center}
\caption{ Results for $\alphazero_\V(8.2~\GeV)$ with $\beta = 6.4$, $6.2$
  and $6.0$ using the $\PS$ splitting. The scale $q_\Box = 3.41/a$;\,
   $q_\Box$ and $a^{-1}$ are in \GeV.}
\label{scaling-table}
\end{table}

\eq{main-result} is our final result. However, to facilitate comparison
with other determinations we convert our result to the ${\rm \overline{MS}}$
scheme, which is related to the $V$ scheme by~\cite{blm}
 \be \label{msb-v}
 \alphanf_\msbar(Q) = \alphanf_\V(e^{5/6}\,Q) \,\left\{ 1 + 2\,\alphanf_\V/\pi
 + \order((\alphanf_\V)^2) \right\}\, .
 \ee
Our result is then equivalent to
 \be
 \alphathree_\msbar(3.56~\GeV) = 0.2203(84) \, ,
 \ee
with the error now dominated by the unknown third-order contribution to
\eq{msb-v}, which we estimate as $(\alphathree_\V)^3 = 0.0075$.

We numerically integrated
the third-order perturbative beta function for
$\alphanf_\msbar$ and applied appropriate matching conditions at
quark thresholds to evolve it to several other scales~\cite{rodrigo}:
 \be
 \alphanf_\msbar(Q) = \cases{
 0.304(17) & \mbox{for $Q=1.7~\GeV\approx M_c$ and $n_f=3,4$}\cr
 0.203(7)  & \mbox{for $Q=5.0~\GeV\approx M_b$ and $n_f=4,5$}\cr
 0.115(2)  & \mbox{for $Q=91.2~\GeV = M_Z$ and $n_f=5$} \, .\cr}
 \ee
The last of these results is consistent with, and significantly
more accurate than, the current world average
$\alphafive_\msbar(M_Z) = 0.117(5)$ .

We believe we have accurately estimated sources of error.
We have checked our result using two
different mass splittings to determine $\ainv$, and four independent
Wilson loops to extract $\alpha_\V(3.4/a)$.  We have checked for
consistency against variations in the quark mass and lattice spacing.
Our estimate of the third-order perturbative contribution to the relation
between $\alpha_\V$ and $\alpha_\msbar$ is consistent with first- and
second-order contributions, and with the third-order terms in
Table~\ref{lnw-table}.
Our result also agrees well with earlier determinations based on lattice
simulations~\cite{aida,latconf-dallas}.  These were
performed without dynamical light quarks.  To estimate the
effect of light quarks, $\alphazero_\V$, computed at $3.41/a$, was run down to
a scale typical of momenta exchanged in the $\Upsilon$ or $\psi$,
determined as in Ref.~\cite{lm,latconf-dallas}.  
There it was equated to $\alphathree_\V$, since it is at this scale
that the two are required to produce the same values for
splittings, and $\alphathree_\V$ was then run back up to the desired scale.
We may repeat
this procedure as an alternative to a direct extrapolation in $n_f$,
now using results for both zero and two light flavors.
This gives $\alphafive_\msbar(M_Z) = .112(4)$ from $n_f=0$
data, and .115(3) from $n_f=2$.  So this method yields values
consistent with, but not as reliable as, extrapolation in $n_f$.

There are prospects for
substantially improving the accuracy of our result fairly soon.
Sources of error in our value for $\alphafive_\msbar(M_Z)$
are listed in Table~\ref{error-table}.  The dominant error
is unrelated to our primary result of \eq{main-result}, but rather
is due to the conversion to ${\rm \overline{MS}}$.
The total error could be cut in half by computing the third-order
correction to \eq{msb-v}, a straightforward perturbative calculation.
The error in $a^{-1}$ will decrease with improved statistics;
we have now completed a new simulation that should
soon reduce it to about .6\%.
Use of an improved gluon action would remove the need for the
$a^2$ correction in the $\PS$ analysis, at little additional
cost~\cite{tasi93}.
Finally, a simulation with either $n_f=3$ or $4$ light quarks would
eliminate the extrapolation error and would require roughly
the same amount of time as for $n_f=2$.
\begin{table}
\begin{center}
\begin{tabular}{lr}
\hline
Source & Uncertainty \\
\hline\\[-.3cm]
Converting from $\alphathree_\V$ to $\alphathree_\msbar$ &  1.7\%  \\[.1cm]
Statistical error in determination of $a^{-1}$           &  .9\%  \\[.1cm]
Extrapolation in $n_f$                                   &  .2\% \\[.1cm]
Finite $a$ and $\order(v^4)$ errors                      &  .2\% \\[.1cm]
Fourth-order evolution of $\alpha_\msbar$                &  .01\% \\[.05cm]
\hline
\end{tabular}
\end{center}
\caption{Sources of error in $\alphafive_\msbar(M_Z)$.}
\label{error-table}
\end{table}

There are a variety of additional lattice calculations that would provide a
broader check on the consistency of our result.
The first would be to repeat our determination using charmonium,
extending previous studies by including dynamical light quarks.
A similar analysis is also possible using light hadrons.
For reasons outlined in the introduction, systematic errors for
light hadrons are not nearly as well understood as for heavy
mesons.  It is important that control of these errors
be improved to demonstrate that these systems yield consistent results.
Finally, simulations which include light-quark vacuum polarization
are still relatively rare.  It would be very useful to repeat this analysis
using an $n_f$ other than two, algorithms for generating gauge-field
configurations other than HMD, and Wilson fermions rather than
staggered.

In this paper we have demonstrated that lattice simulations provide one of
the simplest, most accurate, and most reliable
determinations of the strong coupling
constant. The fact that a lattice simulation of nonperturbative hadronic
structure at scales smaller than 1~GeV agrees with perturbative analyses of
high-energy jet formation is striking confirmation that these diverse
phenomena are governed by a single theory\,---\,QCD. Furthermore, this result
demonstrates our growing mastery over both the nonperturbative and
perturbative aspects of the theory.

An independent lattice determination of $\alphafive_\msbar(M_Z)$
with dynamical light quarks has recently appeared in Ref.~\cite{aoki}.
They obtain results consistent with ours, though with larger errors.

We thank Aida El-Khadra, Paul Mackenzie, Urs Heller and
Jochen Fingberg for several useful discussions;
Greg Kilcup and collaborators, and the HEMCGC collaboration, for generously
providing us with their gauge-field configurations;
and Jonas Berlin for help adapting the HEMCGC configurations for the Cray.
Our simulations were carried out at the Ohio Supercomputer Center
and the National Energy Research Supercomputer Center.
This work was supported by grants from the NSF, the SERC, and the
DOE (DE-FG02-91ER40690, -FC05-85ER250000, -FG05-92ER40742,
-FG05-92ER40722).


\begin{thebibliography}{10}

\bibitem{aida}
A.~X.~El-Khadra, G.~Hockney, A.~S.~Kronfeld, P.~B.~Mackenzie,
Phys.\ Rev.\ Lett.\ {\bf 69}, 729 (1992).  A.~X.~El-Khadra,
Nucl.\ Phys.\ B (Proc.\ Suppl.) {\bf 34}, 141 (1994).

\bibitem{latconf-japan}
G.~P.~Lepage,
Nucl.\ Phys.\ {\bf B} (Proc.\ Suppl.) {\bf 26}, 45 (1992).

\bibitem{onogi}
T.~Onogi, S.~Aoki, M.~Fukugita, S.~Hashimoto, N.~Ishizuka,
H.~Mino, M.~Okawa, A.~Ukawa,
Nucl.\ Phys.\ {\bf B} (Proc.\ Suppl.) {\bf 34}, 492 (1994).

\bibitem{latconf-dallas}
NRQCD Collaboration: G.~P.~Lepage and J.~Sloan,
Nucl.\ Phys.\ {\bf B} (Proc.\ Suppl.) {\bf 34}, 417 (1994).

\bibitem{luscher-schilling}
For approaches which differ from those cited above, see
S.~P.~Booth, D.~S.~Henty, A.~Hulsebos, A.~C.~Irving, C.~Michael and
P.~W.~Stephenson,
Phys.\ Lett.\ {\bf 294B}, 385 (1992);
M.~L\" uscher, R.~Sommer, P.~Weisz and U.~Wolff,
Nucl. Phys. {\bf B413}, 481 (1994);
K.~Schilling and G.~S.~Bali,
Nucl.\ Phys.\ {\bf B} (Proc.\ Suppl.) {\bf 34}, 147 (1994);
and references therein.

\bibitem{blm}
S.~J.~Brodsky, G.~P.~Lepage and P.~B.~Mackenzie, Phys.\ Rev.\
D~{\bf 28}, 228 (1983).

\bibitem{lm}
G.~P.~Lepage and P.~B.~Mackenzie, Phys.\ Rev.\ D~{\bf 48}, 2250 (1993).

\bibitem{worldave}
B.~Webber, 27th International Conference on
High Energy Physics (ICHEP 94), Glasgow, Scotland, July 1994;
I.~Hinchliffe, Meeting of the APS, Division of Particles
and Fields, Albuquerque, New Mexico, August 1994.


\bibitem{gregurs}
Greg Kilcup and his collaborators generously provided us with the
$n_f = 0$ gauge-field configurations,
Urs Heller and his colleagues in the HEMCGC collaboration
with those for $n_f = 2$. The generation of the latter is
described in
K.~Bitar {\it et al.}, Nucl.\ Phys.\ {\bf B {\rm (Proc. Suppl.)} 26}, 259
(1992);
Phys.\ Rev.\ D~{\bf 46}, 2169 (1992).

\bibitem{spect}
C.~T.~H.~Davies, K.~Hornbostel, A.~Langnau, G.~P.~Lepage,
A.~Lidsey, J.~Shigemitsu, and J.~Sloan,
SCRI-94-39, OHSTPY-HEP-T-94-005 (1994),
to appear in Phys.\ Rev.\ D .

\bibitem{mb}
C.~T.~H.~Davies, K.~Hornbostel, A.~Langnau, G.~P.~Lepage,
A.~ Lidsey, C.~J.~Morningstar, J.~Shigemitsu, and J.~Sloan,
SCRI-94-57, OHSTPY-HEP-T-94-004 (1994),
to appear in Phys.\ Rev.\ Lett. .

\bibitem{weisz} P.~Weisz, Phys.\ Lett.\ {\bf 100B}, 331 (1981);
H.~S.~Sharatchandra, H.~J.~Thun and P.~Weisz, Nucl.\ Phys.\ {\bf B192},
205 (1981).

\bibitem{hk} U.~Heller and F.~Karsch, Nucl. Phys. {\bf B251}, 254
(1985); Nucl.\ Phys.\ {\bf B258}, 29 (1985); H.~Hamber and C.~Wu, Phys.\
Lett.\ {\bf 127B}, 119 (1983).

\bibitem{bitar} K.~M.~Bitar {\it et al.}, Phys.\ Rev.\ D~{\bf 48}, 370 (1993).

\bibitem{wilson-loops} The $n_f=0$ Monte Carlo results are from the
Fermilab study discussed in Ref.~\cite{lm}; those for $n_f=2$ are
from the HEMCGC collaboration (Ref.~\cite{gregurs}).

\bibitem{w-pert-th}
The second-order coefficients were obtained by extrapolating those
of Ref.~\cite{hk} to infinite volume; the values for $c_2$ are uncertain
by 1 or 2 in the final digit.
Third-order coefficients are from W.~Dimm, G.~Hockney, G.~P.~Lepage
and P.~B.~Mackenzie, in preparation.  These have a similar uncertainty.

\bibitem{catterall}
The $\beta = 6.2$ results are from
S.~M.~Catterall, F.~R.~Devlin, I.~T.~Drummond and R.~R.~Horgan,
Phys.\ Lett.\  {\bf 321B}, 246 (1994).  The gauge field configurations
for $\beta = 6.4$ were provided by Greg Kilcup and his collaborators.


\bibitem{rodrigo}
For a useful discussion of this procedure, see
G.~Rodrigo and A.~Santamaria, Phys.\ Lett.\ {\bf 313B},
441 (1993).  The quark masses are from Ref.~\cite{mb}.

\bibitem{tasi93}
See, for example,
G.~P.~Lepage, {\em Lattice QCD For Small Computers},
in Proceedings of the 1993 Theoretical Advanced Study Institute,
S.~Raby and T.~Walker, eds.,  World Scientific, Teaneck, N.~J.,
to be published.

\bibitem{aoki}
S.~Aoki, M.~Fukugita, S.~Hashimoto, N.~Ishizuka, H.~Mino,
M.~Okawa, T.~Onogi, and A.~Ukawa,  UTHEP-280 (1994).


\end{thebibliography}
\end{document}